\documentclass[11pt]{article}

\usepackage{color}

\usepackage{greekbf}

\usepackage{amsfonts,amssymb,amsthm}
\usepackage{amsmath,amssymb}
\usepackage{mathrsfs}
\usepackage{enumerate}
\usepackage{srcltx}
\usepackage[ansinew]{inputenc}
\usepackage{mathtools}
    \mathtoolsset{showonlyrefs}
    \mathtoolsset{centercolon}
\usepackage{graphicx}

\DeclareMathAlphabet{\mathbf}{OT1}{cmr}{bx}{it}
\DeclareMathAlphabet{\mathssb}{OT1}{cmss}{bx}{n}
\DeclareMathAlphabet{\mathssn}{OT1}{cmss}{m}{n}
\DeclareMathAlphabet{\mathub}{OT1}{cmr}{b}{n}

\newcommand\remarkname{Remark} %
\newcounter {remarkn}[section]%
\renewcommand \theremarkn {\arabic{section}.\arabic{remarkn}}%
\newenvironment{remark}{
\noindent{\emph{\remarkname}.} \rm%
}

\newcommand\remarksname{Remark} %
\newcounter {remarksn}[section]%
\renewcommand \theremarksn {\thesection.\arabic{remarksn}}%

\begin{document}

\begin{center}

\end{center}
\renewcommand{\a}{\mathbf a}
\renewcommand{\b}{{\mathbf b}}
\newcommand{\Cfil}{{\mathbb C}}
\newcommand{\Sfil}{{\mathbb S}}
\newcommand{\s}{\mathbf s}
\renewcommand{\d}{\mathbf d}
\newcommand{\f}{\mathbf f}
\newcommand{\g}{\mathbf g}
\newcommand{\h}{\mathbf h}
\newcommand{\e}{{\mathbf e}}
\newcommand{\m}{\mathbf m}
\newcommand{\ob}{\mathbf o}
\newcommand{\w}{\mathbf w}
\newcommand{\x}{\mathbf x}
\newcommand{\y}{\mathbf y}
\renewcommand{\P}{{\mathbf P}}
\newcommand{\T}{{\mathbf T}}
\renewcommand{\r}{\mathbf r}
\newcommand{\Q}{{\mathbf Q}}
\newcommand{\G}{{\mathbf G}}
\newcommand{\U}{{\mathbf U}}
\renewcommand{\t}{\mathbf t}
\renewcommand{\u}{\mathbf u}
\renewcommand{\v}{\mathbf v}
\renewcommand{\S}{\mathbf S}
\newcommand{\C}{\mathbf C}
\newcommand{\W}{\mathbf W}
\newcommand{\0}{\mathbf 0}
\newcommand{\F}{\mathbf F}
\newcommand{\N}{\mathbf N}
\newcommand{\A}{\mathbf A}
\newcommand{\V}{\mathbf V}
\renewcommand{\L}{\mathbf \Lambda}

\newcommand{\1}{\mathbf{1}}
\newcommand{\vb}{\mathbf{v}}
\newcommand{\wb}{\mathbf{W}}
\newcommand{\bu}{\mathbf{u}}
\newcommand{\hb}{\mathbf{h}}
\newcommand{\db}{\mathbf{d}}
\newcommand{\Db}{\mathbf{D}}
\newcommand{\ub}{\mathbf{u}}
\newcommand{\nb}{\mathbf{n}}
\newcommand{\mmb}{\mathbf{m}}
\newcommand{\lb}{\mathbf{l}}
\newcommand{\Eb}{\mathbf{E}}
\newcommand{\Hb}{\mathbf{H}}
\newcommand{\Fb}{\mathbf{F}}
\newcommand{\Ib}{\mathbf{I}}
\newcommand{\Xb}{\mathbf{X}}
\newcommand{\xb}{\mathbf{x}}
\newcommand{\gb}{\mathbf{g}}
\newcommand{\Zb}{\mathbf{Z}}
\newcommand{\Yb}{\mathbf{Y}}
\newcommand{\yb}{\mathbf{y}}
\newcommand{\Vb}{\mathbb{V}}
\newcommand{\Mb}{\mathbf{M}}
\newcommand{\rb}{\mathbf{r}}
\newcommand{\tb}{\mathbf{t}}
\newcommand{\ab}{\mathbf{a}}
\newcommand{\Ab}{\mathbf{A}}
\newcommand{\Bb}{\mathbf{B}}
\newcommand{\Cb}{\mathbf{C}}
\newcommand{\Sb}{\mathbf{S}}
\newcommand{\Qb}{\mathbf{Q}}
\newcommand{\Ub}{\mathbf{U}}
\newcommand{\Tb}{\mathbf{T}}
\newcommand{\X}{\mathbf{X}}
\newcommand{\bb}{\mathbf{b}}
\newcommand{\eb}{\mathbf{e}}
\newcommand{\qb}{\mathbf{q}}
\newcommand{\psib}{\mathbf \psi}
\def\cb{\mathbf{c}}
\newcommand{\tr}{\textrm{tr}\,}
\newcommand{\dev}{\textrm{dev}}
\newcommand{\sph}{\textrm{sph}}
\newcommand{\curl}{\textrm{curl}\,}
\newcommand{\Div}{\textrm{Div}\,}
\renewcommand{\div}{\textrm{div}\,}
\newcommand{\sign}{\textrm{sgn}\,}
\newcommand{\Grad}{\textrm{Grad}\,}
\newcommand{\Lin}{\text{Lin}}
\newcommand{\divt}{\text{Div}\,}
%
%

%
%
%

\newcommand\email[1]{\texttt{#1}}
\newcommand\at{:}

\begin{center}
 {\bf \Large
On the Kelvin Problem}
\end{center}
\medskip

\begin{center}
{\large Antonino Favata 
}\end{center}

\begin{center}
 \noindent Dipartimento di Ingegneria Civile, Universit\`a di Roma Tor Vergata\footnote{Via Politecnico 1, 00133 Rome, Italy. \\
{\null} \quad \ Email:
\begin{minipage}[t]{30em}
\email{favata@ing.uniroma2.it} 

\end{minipage}}
\small
\end{center}
\medskip

\begin{abstract}
\noindent {\footnotesize The Kelvin problem of an isotropic elastic space subject to a concentrated  load is solved in a manner that exploits the problem's built-in symmetries so as to determine in the first place the unique balanced and compatible stress field.
\medskip

}
\end{abstract}

\section{Introduction}
This paper deals with a classical problem in linear elasticity, whose solution in terms of displacements was given by Lord Kelvin in a short paper dated 1848 \cite{K}. 
\begin{figure}[h]
\centering
\includegraphics[scale=0.39]{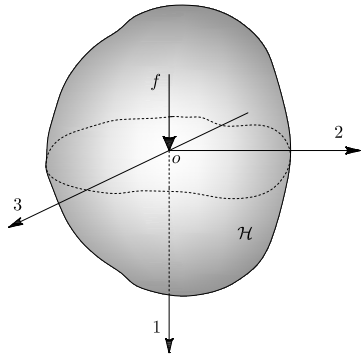}
\caption{The Kelvin problem.}
\label{kelvin3d} 
\end{figure}
The problem (see Figure \ref{kelvin3d}) consists in finding the equilibrium state of a linearly elastic, isotropic material body occupying the whole space and being subject to a point load. Kelvin's is the \textit{fundamental solution} (in modern terminology, the Green function) later used by other XIX century elasticians to determine a number of \textit{strain nuclei} (such as e.g. centers of dilatation and rotation), that is to say, of solutions to problems involving unbounded elastic domains subject to  \emph{force doublets} (or \emph{dipoles}) with or without moment, in various combinations, all inducing singular stress states. 
Singular stress states are also encountered when modeling the strongly localized  residual stress fields due to, say, gas trapped in small cavitites or the thermal mismatch associated with an inclusion. More recently, Kelvin solution has been widely used as a basic ingredient in integral-equation methods for solving boundary-value problems in elastostatics (e.g., in boundary element methods \cite{YX}).

Accounts of Kelvin problem, of different levels of clarity, completeness and purposes, are found in many expositions of the mathematical theory of classic elasticity: we recall the books by  Love \cite{Love}, Sokolnikoff \cite{Sok}, Malvern \cite{Malv}, Benvenuto \cite{Benv}, Davis \& Selvadurai \cite{DS}, Barber \cite{Barb},  and Sadd \cite{sadd}, but there are others.
All these authors tackle the problem by solving the Navier equations for a displacement field having the representation constitutively implied by a tentative representation of  the stress field in terms of a given \textit{scalar potential}. We here propose a different approach, where the basic equations of an elasticity problem --- namely, the equilibrium, constitutive, and compatibility equations --- are not combined preliminarily into one partial differential equation for displacement, as Navier did, but instead used sequentially, in a semi-heuristic fashion, guided by the displacement and stress symmetries intrinsic to the problem at hand.

In the two subsections to follow, we briefly sum up first the classical resolution technique, then ours; certain not completely obvious technical steps of the former, that are not found in any of the elasticity books quoted above, are contained in Appendix 1. Displacement and stress symmetries are discussed in Section 2; the balanced and compatible stress field that solves Kelvin problem is constructed in Section 3, the bulk of this paper; the easy task of deriving the associated strain and displacement fields is carried out in Section 4.

\subsection{Kelvin solution according to Love}\label{subs1.1}
In his paper \cite{K}, Kelvin does not disclose the details of the technique he adopted to solve the problem named after him: in little more than two pages, he just proposes the solution, stressing the similarity with the problem in the theory of thermal conduction, where a  point \textit{heat source} parallels the role of a  \textit{strain source} under form of a concentrated load. We now give a quick account of Kelvin problem and its solution modeled after the account given by Love \cite{Love}. Love's procedure is essentially the same adopted by those authors who solved problems of the same type of Kelvin's after him; while the flow is clear, the mathematical developments are at times skipped. We warn the reader that the notation we use is rather different from the original --- but coherent with the rest of the paper --- and that some slight changes in presentation have been found either necessary or simply convenient.

Navier equation for the displacement vector field $\ub=\ub(x)$ is:
\begin{equation}\label{Navierk}
(\lambda+\mu)\nabla(\Div\ub)+\mu\Delta\ub+\db=\mathbf{0},
\end{equation}
where $\lambda$ and $\mu$ are the Lam\'e constants and $\db$ is the body force.
With a view toward solving this equation, both the displacement and the force field are given a Helmhholtz representation in terms of potential pairs:
\begin{equation}\label{d}
\ub=\nabla\varphi+\curl \w, \qquad   \d=\nabla\psi+\curl\b,  \qquad{\rm with}\;\Div\w=\Div\b=\0.
 \footnote{We recall that, given any sufficiently smooth field $\ub$ over a bounded regular region $\cal R$, its \emph{Helmhholtz representation} consists in  the pair of a scalar field $\varphi$ and a divergenceless vector field $\w$ over $\mathcal{R}$ such that
\begin{equation}
 \ub=\nabla\varphi+\curl \w  \qquad{\rm with}\quad\Div\w=\0,
\end{equation}
(for $\ub\in C(\bar{\cal {R}})\cap C^M(\cal {R})$, $M\geq 1$, both $\varphi$ and $\w$ are of class $C^M(\cal R)$). A straightforward application of the identities $\curl\nabla\varphi=\0$ and $\div\nabla\varphi=\Delta\varphi$, yields: $$\curl\ub=\curl\curl\w\quad\textrm{and}\quad\div\ub=\Delta\varphi.$$} 
\end{equation}
The distance force is taken null in $\mathcal{H}\setminus\mathcal{B}_{\rho}$, where $\mathcal{H}$ is the whole space and $\mathcal{B}_{\rho}\subset{\mathcal H}$ denotes a sphere of radius $\rho$ about the point $o$ where the concentrated load $\f$ is applied.

On recalling that $\Delta\nabla(\cdot)=\nabla\Delta(\cdot)$ and $\Delta\curl(\cdot)=\curl\Delta(\cdot)$, equation \eqref{Navierk} can be written as:
\begin{equation}\label{navhelm}
\nabla\big((\lambda+2\mu)\Delta\varphi+\psi\big)+\curl\big(\mu\,\Delta\w+\nabla\psi+\b\big)=\0.
\end{equation}
A solution of this equation can be obtained by solving the two equations:
\begin{equation}\label{partsol}
(\lambda+2\mu)\Delta\varphi+\psi=0,\qquad
\mu\Delta\w+\b=\0,\footnote{Note that \eqref{navhelm} is recovered by taking the gradient of the first equation and the curl of the second, and then summing up the results.}
\end{equation}
both set over $\mathcal{H}\setminus\mathcal{B}_{\rho}$; the solutions $\varphi_\rho$ and $\w_\rho$  of \eqref{partsol} are:
\begin{equation}
\begin{aligned}
&\psi_\rho(x)=-\frac{1}{4\pi}\int_{\mathcal{B}_\rho}\d(\xi)\cdot\nabla_\xi(\gamma^{-1}(x,\xi))dv(\xi), \\ &\b_\rho(x)=-\frac{1}{4\pi}\int_{\mathcal{B}_\rho}\d(\xi)\times\nabla_\xi(\gamma^{-1}(x,\xi))dv(\xi).
\end{aligned}
 \end{equation}
where, for $\gamma(x,\xi):=|x-\xi|$, $(4\pi\,\gamma)^{-1}$ is the Green kernel of the laplacian. 
The potential pair yielding the representation $\eqref{d}_1$ of the solution to \eqref{Navierk}  is found by taking the limits of $\varphi_\rho$ and $\w_\rho$  for $\rho\rightarrow 0$:
%
%
\begin{equation}\label{phw}
\varphi=\frac{1}{8\pi}\f\cdot \nabla \rho, \quad \w=\frac{1}{8\pi}\,\f\times\nabla \rho, \quad \rho(x):=|x-o|\,.
\end{equation}
%
Finally, the displacement vector is determined by inserting \eqref{phw} in $\eqref{d}_1$. For $\e_2,\e_3$  two unit vectors completing with the direction  $\e_1$  of the applied load  an orthonormal triplet, one finds:
\begin{equation}\label{displLove}
\begin{aligned}
&u_1=-\frac{(\lambda+\mu)f}{8\pi\mu(\lambda+2\mu)}\frac{\partial^2 \rho}{\partial x_1^2}+\frac{f}{4\pi\mu \rho}=\frac{f}{16\pi G(1-\nu)}\left(\frac{2(1-2\nu)}{\rho}+\frac{1}{\rho}+\frac{x_1^2}{\rho^3}\right),\\
&u_2=\frac{(\lambda+\mu)f}{8\pi\mu(\lambda+2\mu)}\frac{\partial^2 \rho}{\partial x_1x_2}=\frac{fx_1x_2}{16\pi G(1-\nu)},\\
&u_3=\frac{(\lambda+\mu)f}{8\pi\mu(\lambda+2\mu)}\frac{\partial^2 \rho}{\partial x_1x_3}=\frac{fx_1x_3}{16\pi G(1-\nu)},
\end{aligned}
\end{equation}
where $u_i:=\u\cdot\e_i$. 

\subsection{Our method}
As shown in the previous subsection, it was Kelvin's concern to find a representation of the displacement field that would make Navier equations \eqref{Navierk} manageable; with his procedure, the elastic state $\{\ub, \Eb, \Sb\}$ is determined sequentially, starting from the knowledge of $\u$. We here propose a different, in fact inverse path. 

Firstly, we find the unique stress field $\S$ in $\mathcal{H}\setminus\{o\}$ that is:
\begin{enumerate}[(i)]
\item \textit{balanced}, in that it satisfies the balance equation
\begin{equation}
\Div\S=\0, 
\end{equation}
\item \textit{compatible}, in that it satisfies the compatibility equation 
\begin{equation}\label{bilita}
\Delta\S+\frac{1}{1+\nu}\,\nabla\nabla(\tr\S)=\0,
\end{equation}
so that, in particular,
\begin{equation}\label{tracciaS}
\Delta(\tr\S)=0;
\end{equation}
\item \emph{balances the applied load}, in the sense that
\begin{equation}\label{eqsurf}
\int_{\partial{\cal N}} \S\nb\,da+\f=\0,
\end{equation}
for every full-dimensional neighborhood ${\cal N}\subset{\mathcal H}$ of the point $o$ where the force $\f$ is applied (here $\nb$ denotes the outer normal to the boundary $\partial{\cal N}$ of ${\cal N}$).
\end{enumerate}

Once the stress field is found, the deformation field $\Eb$ can be computed by using the constitutive equation:
\begin{equation}\label{melat}
\Eb=\frac{1}{2G}\Big (\S - \frac{\nu}{1+\nu}(\tr\S)\Ib \Big ),
\end{equation}
where $G$ and $\nu$ are the shear and Poisson's moduli, and $\Ib$ is the identity tensor. Finally, the determination of the displacement field is accomplished by solving the differential equations:
\begin{equation}\label{eqcomp}
\nabla\u+\nabla\u^T=2\,\Eb.
\end{equation}
\section{Symmetries}
\begin{figure}[h]
\centering
\includegraphics[scale=0.85]{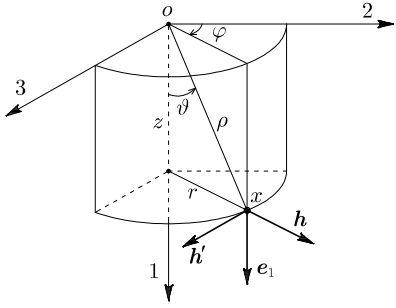}
\caption{Cartesian axes and cylindrical coordinates.}
\label{fig:coordcil} 
\end{figure}

For a problem like Kelvin's, our intuition captures and restitutes a number of symmetries that the displacement and contact-interaction fields must possess. As we shall see, recognizing such symmetries has three direct and useful consequences: 
\begin{enumerate}[(i)]
\item it reduces the representation of the stress tensor to a more tractable form;
\item it makes it simpler to satisfy the compatibility equation;
\item it furnishes a condition that is by all means equivalent to a boundary condition even in the present case of an infinite domain with no boundary.
\end{enumerate}
Different kinds of symmetries can manifest, depending on the body part we consider, either a right circular cylinder whose axis is directed as the load and passes through its application point (Figure \ref{fig:coordcil}) 
or a ball centered at the load. 
\subsection{Symmetries of the displacement field}
We expect the displacement field that solves Kelvin problem to have cylindrical symmetry, that is, when represented in the basis $\{\e,\h,\h^\prime\}$, to be such that both $u_z$ and $u_r$ are independent of $\varphi$ and, in addition, $u_\varphi$ is null:
\begin{equation}\label{simcin}
\u(z,r,\varphi)=u_z(z,r)\e_1+u_r(z,r)\h(\varphi).
\end{equation}
Transforming this expectation into a parametric representation of the candidate solution has two relevant consequences.
\begin{enumerate}
\item \noindent \textit{Consequences on the strain and stress fields.}
In view of the definition of the strain measure $\Eb$, we have that
\begin{equation}
\begin{aligned}
2\,E_{\varphi r}&=u_\varphi,_r+r^{-1}(u_r,_\varphi-u_\varphi),
\\
2\,E_{\varphi z}&=u_\varphi,_z+r^{-1}u_z,_\varphi.
\end{aligned}
\end{equation}
Accordingly, as it is not difficult to see,  whenever \eqref{simcin} prevail, both  strain components $E_{\varphi r}$ and $E_{\varphi z}$ turn out to be everywhere null. But then, given that $\mathcal H$ is supposed to be comprised of an isotropic, linearly elastic material,
both  stress components $S_{\varphi r}$ and $S_{\varphi z}$ must also be everywhere null:
\begin{equation}\label{cylS}
S_{\varphi r}\equiv 0\quad\textrm{and}\quad S_{\varphi z}\equiv 0.
\end{equation}
Therefore, the class of stress fields of interest has the reduced representation:
\begin{equation}\label{bouscil}
\S=\sigma_1\e_1\otimes\e_1+\sigma_2\h\otimes\h+\sigma_3\h^\prime\otimes\h^\prime+\sigma_4(\e_1\otimes\h+\h\otimes\e_1),
\end{equation}
parameterized by the four scalar-valued mappings $\widehat\sigma_i$, with
\begin{equation}\label{bouscilbis}
\begin{aligned}
\widehat\sigma_1(z,r)&=S_{zz}=\S\cdot\e_1\otimes\e_1,\\ \widehat\sigma_2(z,r)&=S_{rr}=\S\cdot\h\otimes\h,\\
\widehat\sigma_3(z,r)&=S_{\varphi\varphi}=\S\cdot\h^\prime\otimes\h^\prime,\\
\widehat\sigma_4(z,r)&=S_{zr}=\S\cdot\e_1\otimes\h.
\end{aligned}
\end{equation}
\item \textit{Consequences on the compatibility condition.}
It follows from \eqref{simcin} that
\begin{equation}
\nabla\u=u_z,_z\e_1\otimes\e_1+u_r,_z\h\otimes\e_1+r^{-1}u_r\h^\prime\otimes\h^\prime+u_z,_r\e_1\otimes\h+u_r,_r\h\otimes\h;
\end{equation}
consequently, 
\begin{equation}\label{definf}
E_{zz}=u_z,_z,\quad
E_{rr}=u_r,_r,\quad
E_{\varphi\varphi}=r^{-1}u_r,\quad
E_{zr}=\frac{1}{2}\Big (u_z,_r+u_r,_z\Big ),
\end{equation}
and hence, in particular, 
\begin{equation}\label{Err}
E_{rr}=(rE_{\varphi\varphi}),_r.
\end{equation}
In view of the inverse constitutive equation \eqref{melat}, this consistency condition of representation \eqref{simcin} can be rewritten as:
\begin{equation}\label{consist}
\sigma_2 -(r\sigma_3),_r+\nu \,r\alpha,_{r}=0.
\end{equation}
\end{enumerate}
In addition to \eqref{simcin}, another symmetry condition prevails, namely,
\begin{equation}\label{lim}
\lim_{r\rightarrow 0^+}u_{r}(z,r)=0;
\end{equation}
as we shall see later on, this last condition works in all respects as a boundary conditions.

\remark Condition \eqref{consist} is related to the compatibility equation 
$$\curl\curl\Eb=\0.$$
In fact,  given the noted symmetries in the displacement and deformation fields, it is not difficult to see that the $\eb_1\otimes\eb_1-$component of $\curl\curl\Eb$ is given by:
\begin{equation}
\curl\curl\Eb\cdot\eb_1\otimes\eb_1=:(\curl\curl\Eb)_{zz}=0=r^{-1}E_{rr,r}-E_{\varphi\varphi,rr}-2r^{-1}E_{\varphi\varphi,r},
\end{equation}
or rather, equivalently,
\begin{equation}
E_{rr,r}=(rE_{\varphi\varphi}),_{rr},
\end{equation}
a direct consequence of \eqref{Err}.

\subsection{Symmetries on the traction}
\begin{figure}[h]
\centering
\includegraphics[scale=0.8]{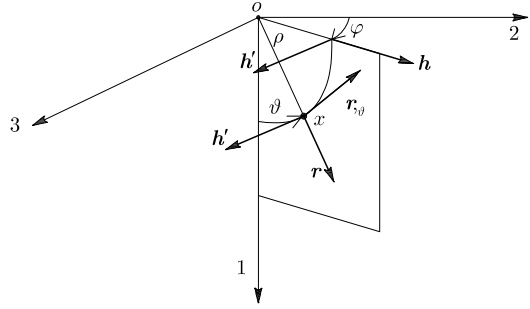}
\caption{Cartesian axes and spherical coordinates.}
\label{fig:coordsf} 
\end{figure}

Recall, to begin with, the following relationships between the cylindrical and spherical coordinates of a given point:
\begin{equation}\label{cilsf}
z=\rho\cos\vartheta,\;\,r=\rho\,|\sin\vartheta|;\quad \rho^2=z^2+r^2,\quad |\tan\vartheta|=\frac{r}{z}
\end{equation}
(Figure \ref{fig:coordsf}).

When written as functions of the spherical coordinates, the stress components $\widehat\sigma_i$ will be denoted by $\widetilde\sigma_i$, with
\begin{equation}\label{rappresent}
\widehat\sigma_i(z,r)=\widehat\sigma_i(\rho\cos\vartheta,\rho|\sin\vartheta|)=:\widetilde\sigma_i(\rho,\vartheta),
\end{equation}
a set of functions assumed to be such that
\begin{equation}\label{rappresent1}
\widetilde\sigma_i(\rho,\vartheta)=\widetilde\sigma_i(\rho,-\vartheta)\;\,\forall \;\rho>0.
\end{equation}
Consider now a body part under form of a ball ${\cal B}_\rho$ centered at the point of application of the load. On writing 
\begin{equation}\label{normal}
\widehat\nb(\rho,\vartheta,\varphi)=\cos\vartheta\e_1+|\sin\vartheta|\,\h(\varphi),\quad (\vartheta,\varphi)\in (0,+\pi)\times (0,2\pi)
\end{equation}
for the outer normal to the boundary $\partial{\cal B}_\rho$ of ${\cal B}_\rho$, the traction on that surface has the expression:
\begin{equation}
\S\nb=(\cos\vartheta\,\sigma_1+|\sin\vartheta|\sigma_4)\e_1+(|\sin\vartheta|\sigma_2+\cos\vartheta\,\sigma_4)\h.
\end{equation}
According to \eqref{eqsurf}, the equilibrium condition is:
\begin{equation}\label{elibri}
\begin{array}{rcl}
f\e_1&=&\displaystyle{-\int_{\partial{\cal B}_\rho}\!\!\S\nb\,da=-\rho^2\int_{0}^{\pi}\Big(\int_0^{2\pi}\S\nb \,d\varphi\Big)|\sin\vartheta|\,d\vartheta}\\&=&\displaystyle{-2\pi\rho^2\Big(\int_{0}^{\pi}\!\!\!(\cos\vartheta\,\widetilde\sigma_1(\rho,\vartheta)+|\sin\vartheta|\widetilde\sigma_4(\rho,\vartheta))|\sin\vartheta|d\vartheta\Big)\e_1},
\end{array}
\end{equation}
whatever $\rho>0$. Therefore, for the right side to remain finite when $\rho$ is chosen arbitrarily big, it is necessary that
\begin{equation}
\begin{array}{rcl}
\widetilde\sigma_1(\rho,\vartheta)&=&\rho^{-2}\widetilde\tau_1(\vartheta)+o(\rho^{-2});\\
\widetilde\sigma_4(\rho,\vartheta)&=&\rho^{-2}\widetilde\tau_4(\vartheta)+o(\rho^{-2}).
\end{array}
\end{equation}
This result suggests the \textit{Ansatz}:
\begin{equation}\label{ansa}
\widetilde\sigma_i(\rho,\vartheta)=\rho^{-2}\widetilde\tau_i(\vartheta),\quad \widetilde\tau_i(\vartheta)= \widetilde\tau_i(-\vartheta)\;\;\,(i=1,\ldots,4),
\end{equation}
which takes into account the parity conditions \eqref{rappresent1}.
\section{The stress field}
%
Given that, when using cylindrical coordinates,
\begin{equation}
\Div\S=(\S\e_1),_z+(\S\h),_r+r^{-1}(\S\h^\prime),_\varphi+r^{-1}\S\h,
\end{equation}
a stress field in the class \eqref{bouscil} is balanced for null distance forces if
\begin{eqnarray}
\0&=&r(\sigma_1\e_1+\sigma_4\h),_z+\big(r(\sigma_2\h+\sigma_4\e_1)\big ),_r+(\sigma_3\h^\prime),_\varphi\nonumber\\&=&\big (r\sigma_1,_z+(r\sigma_4),_r\big)\e_1+\big (r\sigma_4,_z+(r\sigma_2),_r-\sigma_3\big)\h,\nonumber
\end{eqnarray}
in $\mathcal{H}\setminus\{0\}$. The choice of the parameter mappings is therefore restricted to those satisfying the following partial differential equations:
\begin{align}
r\sigma_1,_z+(r\sigma_4),_r&=0,\label{cileq1}
\\r\sigma_4,_z+(r\sigma_2),_r-\sigma_3&=0.\label{cileq2}
\end{align}
%
Moreover, it is not difficult to see that, under the present circumstances, the vectorial compatibility condition \eqref{bilita} is equivalent to a system of four scalar equations, namely,
\begin{equation}\label{compaprima}
\begin{array}{rcl}
\displaystyle{\Delta\sigma_1+\alpha,_{zz}}&=&0,\\\displaystyle{\Delta\sigma_2-2r^{-2}(\sigma_2-\sigma_3)+\alpha,_{rr}}&=&0,\\
\displaystyle{\Delta\sigma_3+2r^{-2}(\sigma_2-\sigma_3)+r^{-1}\alpha,_r}&=&0,\\
\displaystyle{\Delta\sigma_4-r^{-2}\sigma_4+\alpha,_{zr}}&=&0,
\end{array}
\end{equation}
where
\begin{equation}\label{alfatr}
\alpha:=(1+\nu)^{-1}\tr\S=(1+\nu)^{-1}(\sigma_1+\sigma_2+\sigma_3)
\end{equation}
must be harmonic:
\begin{equation}\label{laplace}
  \Delta\alpha=0,
  \end{equation}
to satisfy \eqref{tracciaS}.
To solve \eqref{compaprima}, we  propose the following sequential procedures:
\begin{itemize}
  \item [1.] to determine, a multiplicative constant apart, an appropriate solution $\alpha=\widehat\alpha(z,r)$ of the Laplace equation \eqref{laplace};
  \item [2.] to integrate $\eqref{compaprima}_1$ for $\sigma_1$, in the form:
  \begin{equation}\label{sigmauno}
  \sigma_1=-\Delta^{-1}[\alpha,_{zz}]+c_1\alpha,
  \end{equation}
where $\Delta^{-1}$ denotes the integral operator that formally inverts the laplacian, and $c_1$ is a constant to be determined.
\item [3.] to integrate $\eqref{compaprima}_4$ for $\sigma_4$;
  \item [4.] to determine the fields $\widehat\sigma_2$ and $\widehat\sigma_3$, by solving the system of $\eqref{compaprima}_2$ and \eqref{consist}:
\begin{equation}\label{compaseconda}
\begin{aligned}
&\sigma_2+\sigma_3=\Delta^{-1}[\alpha,_{zz}]+c_2\alpha,\\
&\sigma_2 -(r\sigma_3),_r+\nu \,r\alpha,_{r}=0,
\end{aligned}
\end{equation}
where  $c_2$ is a second constant, such that
\begin{equation}\label{c1c2}
c_1+c_2=1+\nu.\footnote{This condition is arrived at by adding \eqref{sigmauno} and $\eqref{compaseconda}_1$ and by taking into account of \eqref{alfatr}.}
\end{equation}

\end{itemize}

\remark For an alternative procedure, note the following consequence of \eqref{cileq1} and \eqref{cileq2}:
\begin{equation}\label{gradsigma4}
\nabla(r\sigma_4)=-((r\sigma_2),_r-\sigma_3)\e_1-r\sigma_1,_z\h.
\end{equation}
Were step 4. taken right after steps 1. and 2., the field $\widehat\sigma_4$ could be determined by integrating equation \eqref{gradsigma4} along any
regular curve $\cal C$, arc-length parameterized, with tangent $\t$, and going from a fixed point $x_0$ to the variable point $x$: 
\begin{equation}\label{rsigm}
(r\sigma_4)|_{x_0}^x=-\int_{\cal C}((r\sigma_2),_r-\sigma_3)\e_1+r\sigma_1,_z\h)\cdot\t\,ds.
\end{equation}
Remarkably, given the extreme points, the choice of the joining curve is irrelevant, due to a well known result in the theory of differential forms: for $\mathcal{R}$ open and star-shaped, and for $\boldsymbol{\omega}:=\omega_1\e_1+\omega_2\e_2+\omega_3\e_3$ a vector field of class $ C^1(\mathcal{R})$, the differential form $\omega=\omega_1dx_1+\omega_2dx_2+\omega_3dx_3$  is exact if and only if $\curl\boldsymbol{\omega}=\0$ in $\mathcal{R}$. In our case, the differential form and associated vector field under scrutiny are:
$$
\omega=\big((r\sigma_2),_r-\sigma_3\big)\,dz+r\sigma_1,_z\,dr,\quad
\boldsymbol{\omega}=\big((r\sigma_2),_r-\sigma_3\big)\e_1+r\sigma_1,_z\h;
$$
consequently,
$$
\curl \boldsymbol{\omega}=\big(\sigma_1,_{zz}-((r\sigma_2),_r-\sigma_3),_{r}\big)\h'.
$$
Now, by differentiating \eqref{cileq1} with respect to $z$ and \eqref{cileq2} with respect to $r$, and by subtracting the resulting relations, we obtain:
$$
\sigma_1,_{zz}-((r\sigma_2),_r-\sigma_3),_{r}=0,
$$
which allows us to conclude that $\curl \boldsymbol{\omega}=\0$, and hence that $\omega$ is exact.

\subsection{Determination of $\alpha$}
Relations \eqref{ansa} imply a preliminary representation for $\tr\S$:
\begin{equation}\label{rappresentt}
\tr\S=\rho^{-2}\widetilde\tau(\vartheta),\quad\widetilde\tau(\vartheta)=\widetilde\tau(-\vartheta).
\end{equation}
For such a field to be harmonic:
\begin{equation}
\Delta(\rho^{-2}\widetilde\tau(\vartheta))=0,
\end{equation}
function $\widetilde \tau$ must satisfy the following ordinary differential equation:
\begin{equation}\label{ordeq}
\sin\vartheta\,\tau^{\prime\prime}+\cos\vartheta\,\tau^\prime+2\sin\vartheta\,\tau=0,
\end{equation}
whose only \textit{regular} solution is
\begin{equation}\label{boufla}
\widetilde\tau(\vartheta)=\tau_0\cos\vartheta.
\end{equation}
All in all, we are induced to choose
\begin{equation}\label{prealfa}
\tr\S=\tau_0\,\rho^{-2}\cos\vartheta,
\end{equation}
with $\tau_0$ a constant proportional to be applied load, to be determined later on (a method to arrive at \eqref{prealfa} without using \eqref{ansa} is expounded in Appendix 2). At this point, on recalling \eqref{alfatr}, we have:
\[
\alpha=\alpha_0\,\rho^{-2}\cos\vartheta,\quad\textrm{with}\;\, \alpha_0=(1+\nu)^{-1}\tau_0.
\]

\subsection{Determination of $\sigma_1$}
We start by showing how to solve the equation $\eqref{compaprima}_1$, that we here reproduce:
$$
\Delta\sigma_1+\alpha,_{zz}=0,
$$
with the use of the information we gathered so far:  that the unknown field has the preliminary representation $\eqref{ansa}_1$:
$$
\sigma_1=\widetilde\sigma_1(\rho,\vartheta)=\rho^{-2}\widetilde\tau_1(\vartheta),\quad \widetilde\tau_1(\vartheta)=\widetilde\tau_1(-\vartheta),
$$
and that 
\begin{equation}\label{tildea}
\alpha=\widetilde\alpha(\rho,\vartheta)=\alpha_0\rho^{-2}\cos\vartheta\,.
\end{equation}
The equation we have to solve for $\widetilde\tau_1$ is:
\begin{equation}\label{eqtau1}
\tau_1^{\prime\prime}+\cot\vartheta\,\tau_1^\prime+2\tau_1 +3\alpha_0\cos\vartheta(2\cos^2\vartheta-3\sin^2\vartheta)=0.
\end{equation}
A particular solution  of \eqref{eqtau1} can be obtained by the method of parameter variation:
\begin{equation}
\widetilde\tau_1^{(p)}(\vartheta)=\frac{3}{2}\alpha_0\cos^3\vartheta;
\end{equation}
moreover, the even solutions of the homogeneous equation associated with \eqref{eqtau1} are:
$$
\widetilde\tau_1^{(h)}(\vartheta)=\beta_0\cos\vartheta,\quad \beta_0=\textrm{a constant},
$$
 in conclusion, 
\begin{equation}\label{tilde1}
\widetilde\sigma_1(\rho,\vartheta)=\rho^{-2}\big(\frac{3}{2}\alpha_0\cos^3\vartheta+\beta_0\cos\vartheta\big).
\end{equation}

\subsection{Determination of $\sigma_4$}
We have to solve $\eqref{compaprima}_4$ for a field $\widetilde\sigma_4$ that as the representation $\eqref{ansa}_4$, namely,
$$
\sigma_4=\widetilde\sigma_4(\rho,\vartheta)=\rho^{-2}\widetilde\tau_4(\vartheta),\quad \widetilde\tau_4(\vartheta)=\widetilde\tau_1(-\vartheta),
$$
The function
%
$\widetilde\tau_4$ is determined by the ordinary differential equation
\begin{equation}
\tau_4^{\prime\prime}+\cot\vartheta\,\tau_4^\prime+(1-\cot^2\vartheta)\tau_4 +3\alpha_0|\sin\vartheta|(4\cos^2\vartheta-\sin^2\vartheta)=0,
\end{equation}
whose particular solutions are:
\begin{equation}\label{tilde4}
\widetilde\tau_4^{(p)}(\vartheta)=\frac{3}{2}\alpha_0\cos^2\vartheta\,|\sin\vartheta|,
\end{equation}
while the even solutions of the homogeneous associated equation are:
$$
\widetilde\tau_4^{(h)}(\vartheta)=\gamma_0\,|\sin\vartheta|,\quad \gamma_0=\textrm{a constant}.
$$
We conclude that:
\begin{equation}\label{tilde4}
\widetilde\sigma_4(\rho,\vartheta)=\rho^{-2}\big(\frac{3}{2}\alpha_0\cos^2\vartheta|\sin\vartheta|+\gamma_0|\sin\vartheta|\big).
\end{equation}

As to the constants $\beta_0$ and $\gamma_0$, we note that the fields \eqref{tilde1} and \eqref{tilde4} satisfy  the balance equation $\eqref{cileq1}$ only if
$$
\beta_0=\gamma_0;
$$
with this choice, the balance equation $\eqref{cileq2}$ is also satisfied. Furthermore, since the part-wise balance condition \eqref{elibri} can now be written as
\[
f=-2\pi\int_0^\pi(\cos\vartheta\widetilde\tau_1+|\sin\vartheta|\widetilde\tau_4)|\sin\vartheta|\,d\vartheta,
\]
we find that
\begin{equation}\label{condk1}
\alpha_0+2\beta_0=-\frac{f}{2\pi}\,.
\end{equation}

\remark Were we dealing with another classic problem in linear elasticity, Boussinesq's, the problem of a half space with a concentrated load perpendicular to the boundary, we could employ so far the same solving procedure. At this point, though, we would have had to satisfy the boundary condition, and this requires that $\beta_0=0$; the reader is referred to \cite{PGF} for details.

\remark 
On using cylindrical coordinates, we have that
\begin{equation}\label{sigma}
\begin{aligned}
&\widehat\sigma_1(z,r)=\frac{3}{2}\alpha_0\frac{z^3}{(z^2+r^2)^{5/2}}+\beta_0\frac{z}{(z^2+r^2)^{3/2}},\\
&\widehat\sigma_4(z,r)=\frac{3}{2}\alpha_0\frac{z^2r}{(z^2+r^2)^{5/2}}+\beta_0\frac{r}{(z^2+r^2)^{3/2}},\\
&\widehat\alpha(z,r)=\alpha_0\frac{z}{(z^2+r^2)^{3/2}}.
\end{aligned}
\end{equation}
A consequence of the first and third of these relations is:
\[
\widehat\sigma_1(z,r)=\frac{3}{2}\alpha_0\frac{z^3}{(z^2+r^2)^{5/2}}+\frac{\beta_0}{\alpha_0}\widehat\alpha(z,r),
\]
Since 
\begin{equation}\label{alfazz}
\Delta^{-1}[\alpha,_{zz}]=-\frac{3}{2}\alpha_0\frac{z^3}{(z^2+r^2)^{5/2}},
\end{equation}
a comparison with \eqref{sigmauno} shows that 
\begin{equation}\label{c1}
c_1=\beta_0/\alpha_0.
\end{equation}
%

\subsection{Determination of $\sigma_2$ and $\sigma_3$}
%
%
The functions $\sigma_2$ and $\sigma_3$ we seek are to solve system \eqref{compaseconda}, which, with the use of \eqref{c1c2}, \eqref{alfazz}, and \eqref{c1}, can be rewritten as:
\begin{equation}
\begin{aligned}
&\sigma_2+\sigma_3=-\frac{3}{2}\alpha_0\frac{z^3}{\rho^5}+(\alpha_0(1+\nu)-\beta_0)\frac{z}{\rho^3},\\
&\sigma_2 -(r\sigma_3),_r-3\alpha_0\nu\frac{rz}{\rho^5}=0.
\end{aligned}
\end{equation}
This system can be solved sequentially for $\sigma_3$ and $\sigma_2$; it can be checked that solutions have the form:
\begin{equation}
\begin{aligned}
&\widehat\sigma_3(z,r)=-\big(\alpha_0(1-2\nu)-\beta_0\big)\frac{z}{\rho^3}-\big(\alpha_0(1-2\nu)-2\beta_0\big)\frac{z^3}{2r^2\rho^3}+\frac{g(z)}{r^2} ,\\
&\widehat\sigma_2(z,r)=-\frac{3}{2}\alpha_0\frac{z^3}{\rho^5}+(\alpha_0(1+\nu)-\beta_0)\frac{z}{\rho^3}+\\
&\hspace{1.7cm}+\big(\alpha_0(1-2\nu)-\beta_0\big)\frac{z}{\rho^3}+\big(\alpha_0(1-2\nu)-2\beta_0\big)\frac{z^3}{2r^2\rho^3}-\frac{g(z)}{r^2},
\end{aligned}
\end{equation}
parameterized by an arbitrary function $g(z)$.\footnote{Recall that the constant $\beta_0$ can be determined by means of \eqref{condk1} in terms of $\alpha_0$ and the magnitude of the applied load.}

\subsection{Wrapping up} 
At this point, we have satisfied the equilibrium and compatibility equations, and we have balanced the load. However, we have not yet completely found the solution, because the form of function $g(z)$ and the value of the constant $\alpha_0$ are still wanted: we shall find this information by exploiting the symmetry property \eqref{lim} of the displacement field.

By using the inverse constitutive law $(\ref{melat})_2$, we find that
\begin{equation}
E_{\varphi\varphi}=-\frac{1}{2G}\left(\big(\alpha_0(1-\nu)-\beta_0\big)\frac{z}{\rho^3}+\big(\alpha_0(1-2\nu)-2\beta_0\big)\frac{z^3}{2r^2\rho^3}-\frac{g(z)}{r^2}  \right),
\end{equation}
whence
\begin{equation}\label{ur}
u_r=rE_{\varphi\varphi}=-\frac{1}{2G}\left(\big(\alpha_0(1-\nu)-\beta_0\big)\frac{zr}{\rho^3}+\big(\alpha_0(1-2\nu)-2\beta_0\big)\frac{z^3}{2r\rho^3}-\frac{g(z)}{r}  \right).
\end{equation}
With problems formulated on a domain with no boundary,  an incomplete, no matter if inessential, specification of the solution is to be expected, due to an inevitable deficiency of conditions. This is not the case for the Kelvin problem, where the deficit is covered by the symmetry condition \eqref{lim}, namely, 
\begin{equation}
\lim_{r\rightarrow 0^+}u_{r}(z,r)=0,
\end{equation}
which plays the role of a Dirichlet boundary condition. Inserting \eqref{ur} in it, we deduce that:
\begin{equation}\label{condk2}
\alpha_0(1-2\nu)-2\beta_0=0, \quad g(z)=0.
\end{equation}
The first of these relations, together with \eqref{condk1}, yields the values of the constants $\alpha_0$ and $\beta_0$:
\begin{equation}\label{costanti}
\alpha_0=-\frac{f}{4\pi(1-\nu)}, \qquad \beta_0=-\frac{f(1-2\nu)}{8\pi(1-\nu)}\,.
\end{equation}

We are now in a position to write the expression of the stress field:
\begin{equation}\label{eraora}
\begin{aligned}
&\widehat\sigma_1(z,r)=-\frac{f}{8\pi(1-\nu)}\left(3\frac{z^3}{\rho^5}-(1-2\nu)\frac{z}{\rho^3}\right),\\
&\widehat\sigma_2(z,r)=-\frac{f}{8\pi(1-\nu)}\left(3\frac{zr^2}{\rho^5}+(1-2\nu)\frac{z}{\rho^3}\right),\\
&\widehat\sigma_3(z,r)=\frac{f(1-2\nu)}{8\pi(1-\nu)}\frac{z}{\rho^3},\\
&\widehat\sigma_4(z,r)=-\frac{f}{8\pi(1-\nu)}\left(3\frac{z^2r}{\rho^5}-(1-2\nu)\frac{z}{\rho^3}\right).
\end{aligned}
\end{equation}

\remark In case the procedure delineated in Remark 2 is applied, the discovery of condition \eqref{condk1} is delayed. It is convenient to choose the curve $\cal C$ as the union of two different curves $\mathcal{C}=\mathcal{C}_1\cup\mathcal{C}_2$:
\begin{equation}
\mathcal{C}_1:=\{x(z,r,\varphi)\in\mathcal{H}\,|\, z\geq z_0, r=r_0,\varphi=\varphi_0 \},\quad \mathcal{C}_2:=\{x(z,r,\varphi)\in\mathcal{H}\,|\,r\geq r_0,\varphi=\varphi_0\},
\end{equation}
whose tangent vectors are, respectively,
$$
\t_1=\e_1, \quad \t_1=\h(\varphi_0).
$$
The integral in \eqref{rsigm} can be determined as:
\begin{equation}
\int_{\mathcal{C}}\boldsymbol{\omega}\cdot\t=\int_{\mathcal{C}}(\omega_z\e_1+\omega_r\h)\cdot\t=\int_{z_0}^z\omega_z(t,r_0)dt+\int_{r_0}^r\omega_r(z,t)dt,
\end{equation}
with
$$
\omega_z:=(r\sigma_2),_r-\sigma_3  \qquad \omega_r:=r\sigma_1,_z.
$$
It is not difficult to see that
$$
\int_{\mathcal{C}}\boldsymbol{\omega}\cdot\t=\frac{f}{8\pi(1-\nu)}\left(3\frac{z^2r^2}{\rho^5}-(1-2\nu)\frac{zr}{\rho^3}-\frac{(1-2\nu)r_0^4-2(2-\nu)z_0^2r_0^2}{\rho_0^5}\right),
$$
where $\rho_0:=\widehat\rho(z_0,r_0)$. On invoking equation \eqref{rsigm}, we  have that
$$
r\widehat{\sigma}_4(z,r)-r_0\widehat{\sigma}_4(z_0,r_0)=-\frac{f}{8\pi(1-\nu)}\left(3\frac{z^2r^2}{\rho^5}-(1-2\nu)\frac{zr}{\rho^3}-\frac{(1-2\nu)r_0^4-2(2-\nu)z_0^2r_0^2}{\rho_0^5}\right),
$$
whence, by choosing $(z_0,r_0)=(0,0)$, we recover $\eqref{eraora}_4$.

\section{The strain and displacement fields}
To deduce the Kelvin strain field, we have recourse to the inverse constitutive equation
\eqref{melat}. We find:
\begin{equation}\label{straindisp}
\begin{aligned}
&E_{zz}=-\frac{f}{16\pi G(1-\nu)\rho^5}\big(4(1+\nu)z^3+(1-4\nu)zr^2\big),\\
&E_{rr}=\frac{f}{16\pi G(1-\nu)\rho^5}\big(z^3-2zr^2\big),\\
&E_{\varphi\varphi}=\frac{f}{16\pi G(1-\nu)}\frac{z}{\rho^3},\\
&E_{zr}=-\frac{f}{16\pi G(1-\nu)\rho^5}\big(2(2-\nu)z^2r+(1-2\nu)r^3\big).
\end{aligned}
\end{equation}
As to the displacement field, given  \eqref{ur} and \eqref{costanti}, we have that 
\begin{equation}\label{urr}
u_r=\frac{f}{16\pi G(1-\nu)}\frac{zr}{\rho^3};
\end{equation}
moreover, on integrating the obvious combination of  $\eqref{definf}_1$ and \eqref{straindisp}, we 
obtain the following preliminary representation of $u_z$:
\begin{equation}
u_z=\frac{f}{16\pi G(1-\nu)}\left(\frac{2(1-2\nu)}{\rho}+\frac{1}{\rho}+\frac{z^2}{\rho^3}\right)+h(r).
\end{equation}
With this, $\eqref{definf}_4$, and \eqref{urr}, we find that
$$
h^\prime(r)=0\qquad\Leftrightarrow\qquad \h(r)=h_0\e_1,
$$
and we take $h_0=0$ (cf. $\eqref{displLove}_1$). 

\remark
For $h_0\neq 0$, the vector $\hb_0$ represents an arbitrary \textit{translation}  of the whole space in the vertical direction. Such an indeterminacy was to be expected, because the solution of an elasticity problem on a region without boundary is always determined to within a rigid displacement. In this specific case, a vertical translation of an arbitrary amount is the only type of rigid displacement compatible with the postulated symmetries. We regard disposing of such an arbitrariness by setting $h_0=0$ as a covenient \textit{completion of the boundary conditions}.

\section*{Appendix 1. Developments complementing those in Subsection \ref{subs1.1}}
From the classic point of view, a concentrated force is the limit of a body force field having a shrinking support; according to the precise definition found in  \cite{TS}, a sequence $\{\db_n\}$ of body force fields defined on an open neighborhood $\cal{R}$ of a point $o$ tends to the load $\f$ concentrated at $o$ if:
\begin{enumerate}[(i)]
\item $\db_n\in C^2(\cal R)$;
\item $\db_n=\mathbf{0}$ su $\cal R\setminus$$\mathcal{B}_{r_n}(o)$, where $\{\mathcal{B}_{r_n}(o)\}$ is a sequence of spheres of radius  $r_n$ such that  $r_n\rightarrow 0$ when $n\rightarrow\infty$;
\item $\displaystyle{\lim_{n\rightarrow\infty}\int_{\cal R}\db_n=\f}$;
\item the sequence $\{\int_{\cal R}|\db_n|\}$ is bounded.
\end{enumerate}
Given the vector $\d$, the determination of the fields $\psi$ and $\b$ satisfying the Helhmoltz's decomposition $\eqref{d}_2$ can be achieved in two steps.
\begin{enumerate}
\item On applying the divergence operator to the equation \eqref{d}, we obtain the Poisson equation
\begin{equation}\label{poiss}
\begin{aligned}
\Delta \psi=\Div\d, 
\end{aligned}
\end{equation}
whose solution has the well-known representation  (see e.g. \cite{Ev}):
 \begin{equation}
 \psi(x)=-\int_{\mathcal{B}_r}G(x,\xi)\,\Div \d(\xi)\,dv(\xi), \qquad x\in\cal H,
 \end{equation}
where 
\begin{equation}
G(x,\xi)=\big(4\pi\rho(x,\xi)\big)^{-1}, \qquad\rho(x,\xi):=|x-\xi|\,.
\end{equation}
On recalling the identity
$$
\Div(\varphi\v)=\varphi\Div\v+\v\cdot\nabla\varphi,
$$
for $\varphi$ and $\v$ smooth scalar-valued and vector-valued fields, and on using the divergence theorem, we obtain:
\begin{equation}
\psi_r(x)=-\frac{1}{4\pi}\int_{\mathcal{B}_r}\d(\xi)\cdot\nabla_\xi(\rho^{-1}(x,\xi))\,dv(\xi).
\end{equation}
\item On taking the $\curl$ of $\d$, we obtain:
$$
\curl\d=\curl\curl\b=\nabla(\Div\b)-\Delta\b=-\Delta\b.
$$
Thus, we have to solve another Poisson equation:
$$
-\Delta\b=\curl\d,
$$
whose solution is:
$$
\b(x)=\frac{1}{4\pi}\int_{\mathcal{B}_r}G(x,\xi)\curl\d(\xi)\,dv(\xi).
$$
An application of Stokes theorem combined with the identity
$$
\curl(\varphi\v)=\varphi\curl\v+\nabla\varphi\times\vb
$$
yields:
\begin{equation}
\b_r(x)=-\frac{1}{4\pi}\int_{\mathcal{B}_r}\d(\xi)\times\nabla_\xi(\rho^{-1}(x,\xi))dv(\xi).
\end{equation}
\end{enumerate}

We now compute the limits of $\psi_r$ and $\b_r$ for $r\rightarrow 0$, under the assumption that
\begin{equation}
\lim_{r\rightarrow 0}\int_{\mathcal{B}_r}\d(x)\,dv(x)=\f.
\end{equation}
%
We find:
\begin{equation}
\begin{aligned}
\psi(x)=-\frac{1}{4\pi}\f\cdot\nabla(r^{-1}), \quad \b(x)=-\frac{1}{4\pi}\f\times\nabla(r^{-1}), \quad r(x):=|x-o|\,,
\end{aligned}
\end{equation}
or rather, since
$\;
2\nabla(r^{-1})=\Delta\nabla r$,
\begin{equation}
\psi(x)=-\frac{1}{8\pi}\Delta\big(\f\cdot\nabla r\big), \quad \b(x)=-\frac{1}{8\pi}\Delta\big(\f\times\nabla r\big).
\end{equation}
We can now write system \eqref{partsol} as follows:
\begin{equation}
\begin{aligned}
&\Delta\left(\varphi- \frac{1}{8\pi}\f\cdot\nabla r\right)=0,\\
&\Delta\left(\w -\frac{1}{8\pi}\f\times\nabla r \right)=\0,
\end{aligned}
\end{equation}
and read out the particular solution \eqref{phw}.

\section*{Appendix 2. An alternative way to obtain \eqref{prealfa}}
A way to find solutions independent of $\varphi$ to the Laplace equation in cylindrical coordinates:
\begin{equation}\label{cillap}
\Delta\alpha=\alpha,_{zz}+\alpha,_{rr}+r^{-1}\alpha,_r=0
\end{equation}
consists in looking for solutions, if any, having the form
\begin{equation}\label{potvarie}
\widehat\alpha(z,r)=\rho^a z^b r^c.
\end{equation}
It is not dfficult to see that, for such an Ansatz to be successful, the exponents $a,b,c$ must satisfy
%
the following condition:
%
\begin{equation}\label{arm}
a(a+2b+2c+1)+b(b-1)\rho^{2}z^{-2}+c^2\rho^{2} r^{-2} =0\quad\forall z,r>0,
\end{equation}
or rather, equivalently, they must be chosen so as to satisfy the  following three algebraic conditions:
\begin{equation}
c=0,\quad b(b-1)=0,\quad a(a+2b+2c+1)=0.
\end{equation}
There are four possibilities: (i) $b=0$ and $a=0$; (ii) $b=0$ and $a=1$; (iii) $b=1$ and $a=0$; (iv) $b=1$ and $a=-3$; accordingly, 
%
 the desired field must have the following form:
\begin{equation}\label{rapalfa}
\widehat{a}(z,r)=\alpha_0\frac{z}{\rho^3}+\alpha_1 \frac{1}{\rho}+\alpha_2 z+\alpha_3.
\end{equation}

This result can be applied when, as it happens when dealing with Kelvin problem, the trace of a stress field that solves Kelvin problem is found to be proportional to a harmonic field that, when expressed in cylindrical coordinates, does not depend on $\varphi$. In such an application, one starts from the representation \eqref{rapalfa} and quickly sets both constants $\alpha_2$ and $\alpha_3$ to zero, so as to comply with the physical palusibility requirement  that the stress field -- and hence its trace -- vanishes at infinity. Next, one does the same for $\alpha_1$, this time on the basis of an application of a result due to A. Signorini, that we now recall in a version appropriate to our present context (cf. e.g. Sect. 18 of \cite{LTE}). 
\vskip 6pt

\noindent\emph{Signorini's Lemma}. Let $\S$ be a stress field that balances the distance and contact forces $\d$ and $\cb$ acting on a domain $\cal R$ with boundary $\partial \cal R$:
\begin{equation}
\div\S+\d=\0\quad\textrm{in}\;\,{\cal R},\qquad \S\nb=\cb\quad\textrm{on}\:\,\partial{\cal R}.
\end{equation}
Moreover, let $\w$ be a conveniently smooth vector field on $\cal{R}\cup\partial\cal R$. Then,
\begin{equation}\label{signorini}
\int_{\cal{R}}(\nabla\w)\S\,dv=\int_{\cal{R}}\w\otimes\d\,dv+\int_{\partial\cal{R}}\w\otimes\cb\,da.
\end{equation}

\vskip 6pt

\noindent We specialize \eqref{signorini} for $\cal R\equiv\mathcal{B}_\rho$, a ball of radius $\rho$ centered at the point of application of the load, $\d\equiv\0$, and $\w=\mathbf{x}$; by taking the trace of the resulting identity, we obtain:
\begin{equation}\label{signo}
\int_{\mathcal{B}_\rho}\tr\S\,dv=\int_{\partial\mathcal{B}_\rho}\rho\nb\cdot\S\nb\,da.
\end{equation}
Now,
$$
\int_{\partial\mathcal{B}_\rho}\rho\S\nb\cdot\nb\,da=O(\rho),
$$
because we have from \eqref{eqsurf} that
$$
\int_{\partial\mathcal{B}_\rho}\S\nb\,da=O(1).
$$
On the other hand, as to the left side of \eqref{signo}, we have that
\begin{equation}\label{alpha1}
\int_{\mathcal{B}_\rho}\tr\S\,dv=2\pi(1+\nu)\int_0^\rho\int_{0}^\pi(\alpha_0\cos\vartheta+s\alpha_1)\,d\vartheta\,ds;
\end{equation}
for it to be $O(\rho)$ as well, $\alpha_1$ must be set equal to $0$.
%


{\color{black}

\section*{Acknowledgements}
The author gratefully acknowledges the valuable suggestions he received from Prof. Paolo Podio-Guidugli  in the course of a number of extensive discussions.

}

\end{document}